\documentstyle[aps, prl, twocolumn]{revtex}

\newcommand{\beq}{\begin{equation}}
\newcommand{\eeq}{\end{equation}}
\newcommand{\beqa}{\begin{eqnarray}}
\newcommand{\eeqa}{\end{eqnarray}}

\newtheorem{lemma}{Lemma}
\newtheorem{theorem}{Theorem}

\def\ket#1{\left| #1 \right\rangle}

\def\sq{{\textstyle{{1 \over \sqrt 2}}}}
\def\ni{{\noindent}}
\def\01{\{0,1\}}

\begin{document}
\draft
\title{
Classical simulation of quantum entanglement without local hidden variables
}

\author{ Serge Massar$^1$, Dave Bacon$^2$,
Nicolas Cerf$^{3,4}$ and Richard Cleve$^5$\\
{
{$^1$\protect\small\em
Service de Physique Th\'eorique, Universit\'e
Libre de Bruxelles, CP 225,
1050 Brussels, Belgium} \\
$^2$\protect\small\em
Departments of Physics and Chemistry,
University of California Berkeley,
CA 94704, USA }\\
{$^3$\protect\small\em
Ecole Polytechnique, CP 165, Universit\'e Libre de Bruxelles,
B-1050 Bruxelles, Belgium}\\
{$^4$ \protect\small\em
Information and Computing Technologies Research Section,
Jet Propulsion Laboratory,\\
California Institute of Technology, Pasadena, CA 91109, USA}\\
{$^5$\protect\small\em Department of Computer Science, University of
  Calgary, Calgary,
  Alberta, Canada T2N 1N4}}

\date{\today}
\maketitle

\begin{abstract}

Recent work has extended Bell's theorem by quantifying the amount of
communication required to simulate entangled quantum systems with
classical information.
The general scenario is that a bipartite measurement is given from
a set of possibilities and the goal is to find a classical scheme that
reproduces exactly the
correlations that arise when an actual quantum system is measured.
Previous results have shown that, using local hidden variables, a
finite amount of communication suffices to simulate the correlations
for a Bell state.
We extend this in a number of ways.
First, we show that, when the communication is merely required
to be finite {\em on average}, Bell states can be simulated
{\em without} any local hidden variables.
More generally, we show that
arbitrary positive operator valued measurements on
systems of $n$ Bell states can be simulated with $O(n 2^n)$ bits
of communication on average (again, without local hidden variables).
On the other hand, when the communication is required to be
{\em absolutely bounded\/}, we show that a
finite number of bits of local hidden variables is insufficent to simulate
a Bell state.
This latter result is based on an analysis of the non-deterministic
communication complexity of the NOT-EQUAL function, which is constant
in the quantum model and logarithmic in the classical model.

\end{abstract}
\vspace{+0.7cm}

\section{introduction}

We consider how much classical communication is required to simulate
the correlations exhibited by measuring entangled quantum systems.
Following \cite{BCT99}, define a {\em quantum measurement scenario}
as a triple of the form $(\ket{\Psi}_{AB}, M_A, M_B)$, where
$\ket{\Psi}_{AB}$ is an entangled bipartite quantum state, $M_A$ is a
set of measurements on the first component, and $M_B$ is a set of
measurements on the second component.
The goal is to devise communication protocols that enable two separated
parties, Alice and Bob, to simulate a quantum measurement scenario
using classical information.
The {\em input\/} to the protocol is $(x,y) \in M_A \times M_B$, and
Alice receives $x$ (but not $y$) while Bob receives $y$ (but not $x$).
Alice and Bob's {\em outputs\/} should be jointly distributed so as
to exactly reproduce the probability distribution that arises if
an actual quantum system in state $\ket{\Psi}_{AB}$ is measured
according to $(x,y)$.
We shall refer to this problem as {\em classical entanglement simulation}.
In \cite{CCGM}, a related problem, dubbed {\em classical teleportation},
is also introduced.
Here, Alice is given a classical description of a quantum state
$\ket{\Psi}$ and Bob is given a classical description of a quantum
measurement $x \in M$. The goal is for Bob to produce data that stochastically
simulates the result of applying measurement $x$ to state $\ket{\Psi}$.
As shown in \cite{CCGM} and discussed below,
this problem is closely related to classical
entanglement simulation.

The first relevant result in this topic is Bell's famous theorem \cite{Bell},
which implies that, when $\ket{\Psi}_{AB}$ is a Bell state,
there exist $(M_A,M_B)$ for which Alice and Bob must perform {\em some\/}
(non-zero) communication in order to achieve classical entanglement simulation.
More recently, Brassard, Cleve, and Tapp \cite{BCT99},
and independently Steiner \cite{Steiner99}
have shown that, when $\ket{\Psi}_{AB}$ is a Bell state and $M_A$, $M_B$
are each the set of all von Neumann measurements on a qubit, the
simulation is possible with only a {\em finite}
amount of classical communication between Alice and Bob.

In the protocols devised in
\cite{BCT99} and \cite{Steiner99}, it is supposed that Alice
and Bob have an infinite supply of correlated random bits (specifying
real-valued parameters).
Such shared random bits are generally called {\em local hidden
variables}.
The two papers differ in their technical definition of the ``finite amount
of classical communication''.
In \cite{BCT99}, the amount of communication that occurs in the protocol
is {\em exactly\/} 8 bits.
In contrast, (a slightly generalized version of) the protocol in
\cite{Steiner99} has the property that,
for any given pair of measurements $(x,y) \in M_A \times M_B$,
the {\em average} (i.e. expected) number of
bits of communication is 2.97 bits; however, the amount of communication
for any particular execution of the protocol may be arbitrarily large.
The result in \cite{Steiner99} is then refined in \cite{CCGM}, where
the amount of classical communication is decreased to 1.19 bits on
average for all von Neumann measurements.
Also, the sets $M_A$ and $M_B$ are extended to include
all positive-operator-valued measurements (POVMs), using 6.38  bits
of communication on average.
We will refer
to the first kind of protocol as a {\em bounded communication model},
whereas the second kind will be called an {\em average communication
model}.

Regarding the classical entanglement simulation of more than one Bell
state, it is shown in \cite{BCT99} that the exact simulation of arbitrary
von Neumann measurements on $n$ Bell states requires $\Omega(2^n)$ bits
of communication in the bounded communication model. With minor modifications
to the techniques in \cite{BCT99,BCW98}, this $\Omega(2^n)$ lower bound
also carries over to the average communication model.
Also note that this result (as well as most other results for
classical simulation of entanglement) immediately applies
to classical teleportation protocols.
This is because any protocol for classical teleportation of an $n$ qubit state
can be converted into one for classical entanglement simulation of $n$ Bell
states with the same amount of communication.
This is accomplished by Alice first simulating (by herself) the probabilistic
effect of measuring ``her'' $n$ qubits of the $n$ Bell states.
She also computes the
resulting mixture of pure states that describes ``Bob's'' $n$ qubits.
Then Alice classically teleports the state of Bob's $n$ qubits to him.
Conversely, protocols for the classical entanglement simulation can be
converted into protocols for classical teleportation,
at the expense of a little
more communication (see \cite{CCGM} for details).

The present paper generalizes the above results in a number of ways.
All protocols for classical entanglement simulation proposed so far apply to
single Bell states, and they use an infinite number of bits of local
hidden variables for the simulation.
Our first result is that local hidden variables are {\em not} necessary in the
average communication model.
In particular, when $\ket{\Psi}_{AB}$ is a Bell state and $M_A$, $M_B$ are
each the set of all von Neumann measurements, classical entanglement
simulation is possible with a constant number (less than 20) of bits of
communication on average, without any local hidden variables.
We also show
that, when $\ket{\Psi}_{AB}$ consists of $n$ Bell states and $M_A$, $M_B$
are each the set of all POVMs, the simulation can be carried out with no local
hidden variables and $O(n 2^n)$ bits of communication on average.
Note that this communication cost is almost optimal, due to the
aforementioned lower bound of $\Omega(2^n)$.

In contrast to the above results about the the average communication model,
we show that local hidden variables are {\em necessary} in
the bounded communication model (when $\ket{\Psi}_{AB}$ is a Bell state,
and $M_A,M_B$ are all von Neumann measurements).
More precisely, the simulation of $(\ket{\Psi}_{AB},M_A,M_B)$ in the
bounded communication model
requires an {\em infinite\/} number of bits of local hidden variables.
This follows from a connection between the quantum measurement scenario
and the nondeterministic communication complexity of the {\sc not-equal}
function.
These results indicate that there is a fundamental difference
between the absolutely bounded communication model
and the model of communication with bounded expectation.

\section{The case of a single Bell state}

We begin by considering the case of von Neumann measurements on Bell
states. Our first result, stated in Theorem~1, is actually a special
case of a stronger result given in Theorem~3 (where the bound on the
amount of communication will be decreased from 22 to 20 bits, and where
the measurements can be arbitrary POVMs).
The proof of Theorem~1 uses the same basic approach as that of Theorem~3,
but, since it is considerably simpler, it is presented first.

\begin{theorem}
For the quantum measurement scenario $(\ket{\Psi}_{AB}, M_A, M_B)$, where
$\ket{\Psi}_{AB} = \sq (\ket{00} + \ket{11})$, and where
$M_A$, $M_B$ are each the set of all von Neumann measurements, classical
entanglement simulation is possible without any local hidden variables
with a constant number (less than 22) of bits of communication on average.
\end{theorem}

\ni{\bf Proof:}
We first recall Steiner's original protocol \cite{Steiner99}.
The task of the two parties, Alice and Bob, is to simulate carrying
out measurements on the Bell state $\sq (\ket{00} + \ket{11})$ with
respect to operators $R(x)$ and $R(y)$ ($x, y \in [0,1]$), where
\begin{equation}\label{meas}
R(x) = \pmatrix{ \cos(2 \pi x) & \ \ \ \sin(2 \pi x) \cr
                 \sin(2 \pi x) & -\cos(2 \pi x) }.
\end{equation}

In order to carry out this simulation, Alice and Bob share an
infinite sequence of local hidden variables
$\theta_1, \theta_2, \ldots$, which are uniformly distributed
over the interval $[0,1]$.
In addition, Alice has an infinite set of values $u_1, u_2, \ldots$,
which are also uniformly distributed over the interval $[0,1]$.

In order to simulate a Bell state, Alice and Bob carry out the
following operations:
\begin{enumerate}
\item
Alice finds the smallest value $ k \in \{1,2,\ldots\}$ such that
$u_k \le |\cos(2 \pi (\theta_k-x))|$.
Then Alice sends the value of this $k$ to Bob,
and she outputs the value of $\mbox{sign}(\cos(2 \pi (\theta_k-x)))$.
\item
After Bob receives the index $k$ from Alice, he outputs the value of
$\mbox{sign}(\cos(2 \pi (\theta_k-y)))$.
\end{enumerate}
One can verify that this protocol produces the correct statistics
(namely, that Alice and Bob's outputs are random bits, correlated
so as to be equal with probability $\cos^2(\pi(x-y))\ $),
and that the amount of communication is 1.485 bits on average.

Steiner's protocol enables Alice to effectively generate a random
variable, $\theta$, distributed according to the density function
$p(\theta) = {\pi \over 2}|\cos(2 \pi (\theta - x))|$ and convey this
value to Bob.
Explicitly sending the exact value of $\theta$ requires an
infinite number of bits of communication; the above method uses
local hidden variables to accomplish this with a finite amount
of communication.

In order to circumvent the need for local hidden variables (or an
infinite amount of communication), a different approach is used.
Alice generates $\theta$ herself, according to the density function
$p(\theta) = {\pi \over 2}|\cos(2 \pi (\theta - x))|$.
In most cases, only a few bits of $\theta$ suffice for Bob to be
able to compute the value of $\mbox{sign}(\cos(2 \pi (\theta-y)))$.
So Alice sends Bob only a few bits of $\theta$ at a time and
receives a response from Bob each time as to whether or not the
precision is sufficient.
In the first round, Alice sends Bob the first two significant bits
of $\theta$ (since one bit of precision is never sufficient for Bob).
Then Bob determines whether this information unambiguously determines
the value of $\mbox{sign}(\cos(2 \pi (\theta-y)))$ and indicates the
answer in a bit sent to Alice.
In subsequent rounds, Alice sends one additional bit of precision
of $\theta$ to Bob, until Bob's response indicates that the precision
is sufficient.

To upper bound the expected amount of information communicated,
note that, after each round, Bob has at least a $1 \over 5$ chance of
having $\theta$ with sufficient precision.
This is because, for any $z \in [0,1]$ and $w \in [0,{1 \over 4}]$,
\begin{equation}
\int_{z-w}^z {\textstyle{\pi \over 2}}|\cos(2 \pi \theta)| d\theta
> {\textstyle{1 \over 4}}
\int_z^{z+w} {\textstyle{\pi \over 2}}|\cos(2 \pi \theta)| d\theta.
\end{equation}
Thus, the expected number of rounds is less than 5.
Since the first round consists of 3 bits (two from Alice and one
from Bob) and each subsequent round consists of two bits (one from
each of Alice and Bob), the expected number of bits of communication
is less than 11.
To simulate an arbitrary von Neumann measurement, it suffices to
simulate two measurements with respect to operators of the form $R(x)$
\cite{BCT99}.
Thus, the expected amount of communication is less than 22 bits.
$\Box$\vspace*{2mm}

Regarding the {\em minimum} number of bits of communication
necessary to perform classical entanglement simulation without local
hidden variables as in Theorem 1,
it should be noted that a single run of a protocol without
local hidden variables cannot succeed in general
if the communication is less than one bit.
This is because, in the case where the two
measurements $x$ and $y$ are both in the same basis, Alice's and
Bob's outputs have exactly one bit of mutual information.
One can easily check that this mutual information is a lower bound
on the amount of forward and backward communication that must be used
to simulate entanglement (see Appendix).
Therefore, one bit of communication is necessary in this worst case.
For other specific pairs of measurements ($x$, $y$), the mutual
information is lower than one bit, and so is the minimum
amount of communication.
Let us now consider the case where the measurement directions
$x$ and $y$ are chosen at random, and assumed to be
isotropic (the distribution of maximum uncertainty).
The average communication here is with respect to the probabilistic
selection of a pair of measurements as well as the probabilistic
choices made by Alice and Bob during the execution of the protocol.

\begin{lemma}\label{ll0}
Let $(\ket{\Psi}_{AB}, M_A, M_B)$ be the quantum measurement scenario where
$\ket{\Psi}_{AB} = \sq (\ket{00} + \ket{11})$, and $M_A$, $M_B$ are each
the set of all von Neumann measurements.
Suppose that a pair $(x,y)$ is selected according to 
two independent
uniform
distributions on the surface of the Bloch sphere.
Then, for any protocol in the average communication model that has
no local hidden variables, the sum of the (forward and backward)
communication must be at least 0.279~bits on average.
\end{lemma}

\ni{\bf Proof:}
Consider the situation where Alice and Bob
are each given a random measurement direction ($\vec x$ and $\vec y$)
by a third person, say Charles. As before, Bob does not know the measurement
Alice is performing, and, conversely, Alice ignores Bob's measurement.
We also assume that, initially, the two parties share no information.
Then, Charles observes the outcomes of Alice's and Bob's measurements,
noted $a$ ($=\pm 1$) and $b$ ($=\pm 1$). Our goal here is
to estimate the number of bits that must be communicated
from Alice to Bob (and from Bob to Alice) in order for them to exactly
reproduce the quantum correlations that would be observed if
they shared a singlet. This quantity can be bounded from below
by the amount of shared randomness that Charles observes
between Alice's and Bob's outcomes, while {\em knowing}
the measurement directions. In other words, what is relevant here
is the mutual information between Alice's and Bob's outcomes
$a$ and $b$ {\em conditionally} on the measurement directions $x$ and $y$,
that is, $I(a{\rm:}b|x,y)$. (It is shown in the Appendix that
the mutual information is indeed a lower bound on the number of bits
that must be communicated.)
For given measurement directions $\vec x$ and $\vec y$,
the correlation coefficient is $r=-\vec x \cdot \vec y$,
so that the joint distribution of the outcomes is
$p(a,b|r)=(1+r\, a\, b)/4$, with $a=\pm 1$ and $b=\pm 1$.
The resulting mutual information for a given $r$ is then equal to
\beq
I(a{\rm:}b|r)={1+r\over 2}\log_2(1+r) + {1-r\over 2}\log_2(1-r)
\eeq
If $\vec x$ and $\vec y$ are uniformally distributed, then the correlation
coefficient is distributed as $P(r)=1/2$ in the interval $[-1,1]$.
As a result, the (average) mutual information between $a$ and $b$
conditionally on $r$ can be written as
\beqa  \label{1stcalc}
I &=&\int  I(a{\rm:}b|r) \; P(r)\; dr  \nonumber\\
  &=& \int_{-1}^1 (1+r)\log_2(1+r) \; dr  \nonumber \\
  &=& \log_2(2/\sqrt{\rm e})
\eeqa
Thus, the amount of (forward and backward) communication
that is necessary to establish this shared randomness
between Alice and Bob is bounded by
$C_f+C_b \geq I = 0.279$~bits.
$\Box$\vspace*{2mm}

Note that this bound assumes that there are
no initially shared local hidden variables between Alice and Bob.
In a more general scenario, however, the bound in Lemma 1 only measures
the {\em total} amount of shared randomnes,
possibly including prior shared randomness.
In other words, $I$ is the sum of the initial shared randomness
and the communication, so it does not discriminate
the random bits that are shared beforehand (the local hidden
variables) from the bits that are communicated after the
measurement basis are disclosed to Alice and Bob.

Now, we shall show that, if the bound on the communication is changed from
being constant on average to being an absolute constant, then
the classical entanglement simulation without local hidden variables
that occurs in Theorem 1 becomes impossible to achieve.
Prior to doing this, we review a relevant result from the theory of
communication complexity (see \cite{comcomp} for an extensive review of
the field).
Consider the {\sc not-equal} function,
$\mbox{\it NE} : \01^n \times \01^n \rightarrow \01$, defined as

\begin{equation}
\mbox{\it NE\/}(x,y) = \cases{1 & if $x \neq y$ \cr
                            0 & if $x = y$,\cr}
\end{equation}
and suppose that Alice and Bob are given $x$ and $y$ respectively as
inputs and their goal is to evaluate $\mbox{\it NE\/}(x,y)$ in the following
weak sense.
Bob should output a bit $b$ that is distributed so that:
if $\mbox{\it NE\/}(x,y) = 0$ then $\Pr[b=1] = 0$;
if $\mbox{\it NE\/}(x,y) = 1$
then $\Pr[b = 1] > 0$.
Also, assume that Alice and Bob have no {\em a priori\/} shared random
information.
A protocol that accomplishes this can be regarded as a
{\em nondeterministic\/}
protocol for the {\it NE\/} function.
By standard techniques in communication complexity
(see p.~19, \cite{comcomp}),
the following lower bound can be obtained on the amount of communication
required by Alice and Bob in order to achieve this.

\begin{lemma}\label{ll1}
Any nondeterministic {\em classical\/}
protocol for computing the function {\it NE} requires
at least $\log_2(n)$ bits of communication.
\end{lemma}

The above lemma will be used to prove the following theorem.

\begin{theorem}\label{th3}
For the quantum measurement scenario $(\ket{\Psi}_{AB}, M_A, M_B)$, where
$\ket{\Psi}_{AB} = \sq (\ket{00} + \ket{11})$, and where $M_A$, $M_B$ are
each the set of all von Neumann measurements, classical entanglement
simulation is impossible if the number of bits of communication is absolutely
bounded by a constant and the number of bits of local hidden variables is
also a finite constant.
\end{theorem}

\ni{\bf Proof:}
We will show that any protocol for classical entanglement simulation
that uses a constant number of bits of communication (in the absolute
sense) and a constant number of bits of local hidden variables can
be converted into a nondeterministic protocol for {\it NE\/} with a
constant amount of communication (independent of $n$), thereby
contradicting Lemma~\ref{ll1}.
First, note that a finite amount of prior shared randomness can always be
simulated by a finite amount of communication at the start of the protocol
(to establish the shared randomness).
Thus, we can suppose, without loss of generality, that the protocol
for classical entanglement simulation
uses no prior shared randomness.

Consider the restricted set of measurements, where $M_A$ and $M_B$ each
consist of all measurements with respect to the operators of the form
$R(x / 2^n)$ ($R$ is defined in Eq.~\ref{meas}), where $x \in \01^n$
is an $n$-bit binary number.
Note that if the protocol for entanglement simulation is given
$x$ and $y$ as inputs then the resulting output bits of Alice and Bob,
call them $a$ and $b$, satisfy
$\Pr[a=b] = \cos^2(\pi (x-y)/2^n)$.
It follows that: $\Pr[a \neq b] = 0$ if $x = y$; and $\Pr[a \neq b] > 0$ if
$x \neq y$.
Therefore, if at the end of the protocol Alice sends her bit $a$ to Bob
(increasing the communication cost of the protocol by one bit) and then
Bob outputs $a \oplus b$, the result is a nondeterministic protocol
for computing {\it NE\/} with a constant number of bits of communication
(independent of $n$), contradicting Lemma~\ref{ll1}.
$\Box$\vspace*{2mm}

In contrast to Lemma \ref{ll1}, we note the following.

\begin{lemma}\label{ll2}
There is a nondeterministic {\em quantum\/} protocol for {\it NE}
where the communication cost is exactly one qubit.
\end{lemma}

\ni{\bf Proof:}
The idea is for Alice to create the state
$\cos(\pi x / 2^n)\ket{0} + \sin(\pi x / 2^n)\ket{1}$ and send it to Bob.
Then Bob measures with respect to the operator $R(y)$.
It is straightforward to calculate that the outcome of Bob's measurement $b$
satisfies $\Pr[b = 1] = 0$ if $x = y$, and $\Pr[b = 1] > 0$ if $x \neq y$.
$\Box$\vspace*{2mm}

Comparing Lemmas~\ref{ll1} and~\ref{ll2}, there is a 1 vs.\ $\log_2(n)$
quantum vs. classical gap for the nondeterministic communication complexity
of {\it NE}.
This is noteworthy since it is a case where even an exponential increase
in the amount of communication permitted is not sufficient for a classical
protocol to simulate a quantum protocol.
See \cite{deWolf00} for other results about nondeterministic communication
complexity in the quantum case.

\section{The case of several Bell states}\label{3}

In this section, we shall exhibit a protocol that generalises theorem
1 as follows.

\begin{theorem}\label{th5}
Suppose that Alice and Bob must simulate the classical teleportation
of a state belonging to a $2^n$ dimensional Hilbert space
and suppose that Bob must carry out an arbitrary POVM.
Or suppose that Alice and Bob must simulate carrying out arbitrary
POVM's
on $n$ ebits
(that is an entangled state belonging to the tensor product of two $2^n$
dimensional Hilbert spaces).
Both simulations can be
realized by communicating on average less then
 $(3n + 6)2^{n} + 2$ bits. Specifically we shall exhibit a protocol
 in which
Alice sends Bob on average less then $(3n + 6)2^{n}$
bits and Bob sends Alice on average less then $2$ bits of
communication.
\end{theorem}

We now proceed with a general proof of Theorem~\ref{th5}.
We start with a discussion of how much classical
communication is necessary
for approximate simulation of quantum communication.

\begin{lemma}\label{l3} Consider the problem of classical
teleportation in which Alice is given a quantum state $|\Psi\rangle$
belonging to a $2^n$ dimensional Hilbert space (i.e.\ Alice is given $n$
qubits) and Bob is given an arbitrary POVM $x=\{B_l\}$ where $B_l$ are
the POVM elements.
Suppose Alice sends Bob $(m+1)2^{n+1}$ (with $m \geq n/2$) bits of classical
information about state $|\Psi\rangle$. With this classical
information
Bob can
calculate an approximation $P^m(l)$ to the true probability
$P(l) = \langle \Psi |B_l |\Psi \rangle$
that his measurement yields outcome $l$.
Alice can choose the bits she sends to Bob, and Bob
can use an algorithm, such that the
approximate probabilities sum to 1 ($\sum_l P^m(l) =1$) and satisfy
the constraint
\beq
| P(l) - P^m(l) | \leq \alpha^m Tr (B_l)
\label{PP}
\eeq
where $Tr (B_l)$ is the trace of the POVM element $B_l$ and $\alpha^m$
is bounded  by
\beq
\alpha^m <  2^{n/2 - m +1 } \ .
\label{alpha}
\eeq
An equivalent formulation of (\ref{PP}) is that the information
provided by Alice enables Bob to define two bounds
\beqa
P_{min}^m (l) &=& \max \{ 0, P^m(l) -  \alpha^m   Tr (B_l) \} \ ,
\label{Pmin}
\\
P_{max}^m (l) &=& \min \{ 1, P^m(l) +  \alpha^m   Tr (B_l) \} \
\label{Pmax2}
\eeqa
such that he knows with certainty that $P(l)$ belongs to the interval
\beq
P(l) \in
[P_{min}^m (l) , P_{max}^m (l)] \ .
\label{PPP}
\eeq
 This interval has the property that
as $m$ increases the interval shrinks:
\beqa
0 &\leq& P_{min}^m (l) \leq
P_{min}^{m+1} (l)\ , \label{Pmin1}
\\
1 &\geq& P_{max}^m (l) \geq
P_{max}^{m+1} (l) \ .
\label{PPmin}
\eeqa
\end{lemma}

\ni{\bf Proof:} Let us
choose an arbitrary basis $|j\rangle$ of the Hilbert space.
This basis is known
to both Alice and Bob. In this basis,
the state $|\Psi \rangle$ can be written as
\beq
|\Psi \rangle = \sum_{j=1}^{2^n}
\left( X(j) + i Y(j) \right) |j\rangle
\eeq
where $X(j)$ and $ Y(j)$ are real numbers. We can write them as
\beqa
X(j) &=& (-1)^{x_0(j)}\sum_{r=1}^\infty x_r(j) 2^{-r}\ ,\nonumber\\
Y(j) &=& (-1)^{y_0(j)}\sum_{r=1}^\infty y_r(j) 2^{-r}\ .
\label{coeff}
\eeqa
were $x_r(j) , y_r(j)\in \{0,1\} $.

We shall suppose that the $(m+1)2^{n+1 }$ bits of information about
$|\Psi\rangle$ sent by Alice are the values of
$x_r(j), y_r(j) $ for all $j$
and for $0\leq r \leq m$. Bob then knows
the coefficients $X(j), Y(j)$ with finite precision.
Denote the part of $X(j)$ and $ Y(j)$ which is known to Bob by
\beqa
X^m(j)&=& (-1)^{x_0(j)}\left( \sum_{r=1}^m x_r(j) 2^{-r} + 2^{-m-1} \right)
\ ,\nonumber\\
Y^m(j)&=& (-1)^{y_0(j)}\left( \sum_{r=1}^m y_r(j) 2^{-r} + 2^{-m-1} \right)
\ .\eeqa
We then have
\beqa
|X(j) - X^m(j) | \leq 2^{-m-1} \ ,\nonumber\\
|Y(j) - Y^m(j) | \leq 2^{-m-1}\ .
\eeqa

Denote Bob's estimate of the state $\Psi$ by
\beq
|\Psi^m\rangle =
\sum_{j=1}^{2^n}
\left( X^m(j) + i Y^m(j) \right) |j\rangle \ .
\label{Phim}
\eeq
We can write the true state as
\beq
|\Psi\rangle = |\Psi^m\rangle +
|\Delta \Psi^m\rangle  \ .
\eeq
Bob's uncertainty can by measured by
\beqa
\langle \Delta \Psi^m |\Delta \Psi^m\rangle
&=& \sum_{j=1}^{2^n} (X(j) - X^m(j) )^2 + (Y(j) - Y^m(j)
)^2\nonumber\\
&\leq& 2^{n -2m -1}
\label{Dphi}
\eeqa
For this inequality to be informative,
 it is necessary that $m
\geq n/2$.

Bob's estimate for the probability $P(l)$ is
\beq
P^m(l) =  \langle
\Psi^m | B_l|\Psi^m\rangle \ .
\eeq
Let us write Bob's POVM elements as
\beq
B_l = Tr (B_l) |\tilde \beta_l\rangle \langle \tilde \beta_l |
\eeq
where $|\tilde \beta_l\rangle$ is a normalized state.
We then have
\beqa
P(l) - P^m(l) &=& Tr (B_l) \left [
2 {\rm Re } \langle \Psi^m |\tilde \beta_l\rangle \langle \tilde \beta_l |
\Delta \Psi^m\rangle \right.\nonumber\\
& &
\left.
+
|\langle \tilde \beta_l |
\Delta \Psi^m\rangle |^2 \right ]
\eeqa
which we can bound by
\beqa
& &|P(l) - P^m(l) |\nonumber\\
&\leq& Tr (B_l) \left [
2 | \langle \Psi^m | \tilde\beta_l\rangle|| \langle \tilde \beta_l |
\Delta \Psi^m\rangle |
+
|\langle \tilde \beta_l |
\Delta \Psi^m\rangle |^2 \right ] \nonumber\\
&\leq& Tr (B_l) \left [
2 | \langle \tilde \beta_l |
\Delta \Psi^m\rangle |
+
|\langle \tilde \beta_l |
\Delta \Psi^m\rangle |^2  \right ]\nonumber\\
&\leq& Tr (B_l) \left [
2 \sqrt{ \langle \Delta \Psi^m |\Delta \Psi^m\rangle }
+\langle \Delta \Psi^m |\Delta \Psi^m\rangle \right ]
\nonumber\\
&\leq&  Tr (B_l) \left [ 2^{n/2 -m +1/2} + 2^{n - 2m -1}\right ]\nonumber\\
&\leq& Tr (B_l)  \alpha^m
\eeqa
where we have used that $2^{n/2 -m +1/2} + 2^{n - 2m -1} < \alpha^m$
if $m\geq n/2$.
This proves (\ref{PP})

Let us now prove the monotonicity properties (\ref{Pmin1}) and
(\ref{PPmin}). To this end we compute the difference between
successive estimates $|P^m(l) -
P^{m+1}(l)|$. Define the quantities $\delta X^{m+1}(j)$, $\delta
Y^{m+1}(j)$ by
\beqa
\delta X^{m+1}(j) &=&  X^m(j) - X^{m+1}(j) \ , \nonumber\\
\delta Y^{m+1}(j) &=&  Y^m(j) - Y^{m+1}(j)\ .
\eeqa
We have
\beq
|\delta X^{m+1}(j) |  \leq 2^{-m-2} \quad , \quad
|\delta Y^{m+1}(j) |  \leq 2^{-m-2}
\ .
\eeq
Then we define the
difference of estimated state for 2 successive
values of $m$:
\beqa
|\delta \Psi^m \rangle &=& |\Psi^m\rangle - |\Psi^{m+1}\rangle
\nonumber\\
&=& \sum_{j=1}^{2^n}
\left( \delta X^{m+1}(j) + i \delta Y^{m+1}(j) \right) |j\rangle \ .
\eeqa
The difference between successive estimates decreases as
\beqa
\langle \delta \Psi^{m+1} |\delta \Psi^{m+1}\rangle
&=& \sum_{j=1}^{2^n}  (\delta X^{m+1}(j) )^2 + ( \delta Y^{m+1}(j)
)^2\nonumber\\
&\leq& 2^{n -2m -3} \ .
\eeqa
Finally we have
\beqa
& &|P^m(l) - P^{m+1}(l)|\nonumber\\
 &=& Tr (B_l)
 \left | 2 {\rm Re } \langle \Psi^m | \tilde \beta_l\rangle
\langle \tilde \beta_l |
\delta \Psi^{m+1}\rangle
+
| \langle \tilde \beta_l |
\delta \Psi^{m+1}\rangle |^2 \right | \nonumber\\
&\leq& Tr (B_l) \left [
2 \sqrt{ \langle \delta \Psi^{m+1} |\delta \Psi^{m+1}\rangle }
+\langle \delta \Psi^{m+1} |\delta \Psi^{m+1}\rangle \right ]
\nonumber\\
&\leq&  Tr (B_l) \left [ 2^{n/2 -m -1/2} + 2^{n - 2m -3}\right ]\nonumber\\
&\leq& Tr (B_l)  \alpha^{m+1} \quad \mbox{(if $m \geq n/2 -1$)}
\eeqa
which together with the definitions (\ref{Pmin}) and (\ref{Pmax2})
implies (\ref{Pmin1}) and (\ref{PPmin}).
$\Box$\vspace*{2mm}

We now turn to the proof of Theorem~\ref{th5}.  The two complications
with respect to Theorem~1 are that the state is
described by many parameters and not one angle $x$ and
that Bob may have
more than $2$ outcomes between which to choose since his POVM may have more
than $2$ outcomes.
These two complications lead to the more
intricate protocol given below.
\vspace*{2mm}

\ni{\bf Proof of Theorem \ref{th5}:}
Note that the second part of the theorem (dealing with simulating
measurements on ebits) follows directly from the first part of the
theorem (dealing with the simulating the transmission of qubits)
in view of the relationships between classical teleportation and
classical entanglement simulation (discussed at the end of Section~1).
Hence we consider only the first part dealing with the simulation of
quantum communication.

The protocol used by Alice and Bob consists of a series of rounds
which we label by $K$. Alice's role during each round is simple to
describe. She starts the round by sending Bob some information about
the state $|\Psi\rangle$. Specifically during the first round ($K=1$)
this information consists
 of the values of the coefficients $x_r(j) ,
y_r(j)$ defined in (\ref{coeff})
for $r=0,\ldots, 3n/2 + 2$ and all values of $j$
($j=1,\ldots,2^n$).
During the next rounds
($K=2,3,\ldots$) this information consists of the
values of the coefficients $x_{3n/2 + K+1}(j) ,
y_{3n/2+K+1}(j)$ for $j=1,\ldots,2^n$.

Upon receiving this information, Bob will carry out a computation
(which we describe below) and reaches one of two conclusions. One
possibility is that he is able to choose an outcome $l$ for his
measurement. The second possibility is that he is unable to choose an
outcome $l$ in which case he needs more information about $|\Psi\rangle$.
Thus the end of the round consists of Bob sending Alice one bit
telling her whether or not he needs more information about
$|\Psi\rangle$. If Bob does not need more information, then the
protocol terminates since Bob has chosen an outcome for his
measurement.
If Bob needs more information, then they both increments $K$
by 1 and the next round starts.

The reason why the first round differs slightly from the next rounds
is that Bob needs a large amount of initial information before he can
start trying to choose an outcome. If this first try does not succeed,
then only small additional
amounts of information are necessary for Bob to try again to choose an
outcome. Mathematically the necessity for the large amount of initial
information is expressed in equations (\ref{TM}) and (\ref{muRK}) below
which are non trivial inequalities only when a sufficient amount of
information has been transmitted by Alice to Bob.

We now describe the computation carried out by Bob. Recall
that with the
information sent to him by Alice, Bob can construct the approximation
$|\Psi^{3n/2 +K+1}\rangle$ to the true state $|\Psi \rangle$ (defined
in (\ref{Phim})).
Using this approximate state,
he knows that the true probability $P(l)$ of outcome $l$ is
comprised between $P_{min}^{3n/2 +K+1}(l)$ and $P_{max}^{3n/2 +K+1}(l)$,
see (\ref{PPP}).
It is convenient for Bob to re-express this approximate knowledge of
the true probabilities in terms of set of subintervals $I^K(l) , R^K$
of the unit interval $[0,1[$. Bob's strategy will then be simply
expressed in terms of these intervals.

To define these subintervals we introduce the following
notations
\beq
\Delta^m(l)= P_{min}^{m}(l) - P_{min}^{m-1}(l)
\eeq
(note that $\Delta^m(l) \geq 0$, see (\ref{Pmin1})) and
\beq
T^m = \sum_{l=1}^L P_{min}^{m}(l)
\eeq
where $L$  is the number of
outcomes of Bob's POVM $\{ B_l \}$, $l=1,\ldots, L$.
We have the following property
\beqa
T^m &\geq&  \sum_{l=1}^L P^{m}(l) - \alpha^m Tr (B_l)
\nonumber\\
&=& 1 - \alpha^m \sum_{l=1}^L Tr (B_l)\nonumber\\
&=&  1 - \alpha^m 2^{n}\nonumber\\
&\geq&  1 - 2^{3n/2-m +1} \quad \mbox{(if $m \geq  n/2 +1$)}
\label{TM}
\eeqa
which follow from (\ref{Pmin}) and (\ref{alpha}).

The subintervals are defined for $K=1$ by
\beqa
I^1(1) &=& [0\ ,\ P^{3n/2+2}_{min}(1)[ \ ,\nonumber\\
I^1(l) &=& [\sum_{l'=1}^{l-1}  P^{3n/2+2}_{min}(l')\ ,\
\sum_{l'=1}^{l}  P^{3n/2+2}_{min}(l')[
\ ,
\eeqa
and for $K=2,3,\ldots,\infty$ by
\beqa
I^{K}(1) &=& [T^{3n/2 +K}\ ,\  T^{3n/2 +K} +  \Delta^{3n/2 +K+1}(1)
[\ ,\nonumber\\
I^{K}(l) &=& [T^{3n/2 +K} +
\sum_{l'=1}^{l-1}  \Delta^{3n/2 +K+1}(l')\ ,\nonumber\\
& &\
T^{3n/2 +K} +
\sum_{l'=1}^{l}  \Delta^{3n/2 +K+1}(l')[
\ .\eeqa
We also define the subintervals
\beq
R^{K} = [T^{3n/2 +K+1}\ ,\ 1[ \quad , \quad K = 1,2,\ldots \infty\ .
\eeq

These subintervals have several properties which follow
directly
from
equations (\ref{PP}),(\ref{Pmin}),(\ref{Pmin1}), (\ref{PPP}) and
(\ref{TM}):

\begin{enumerate}

\item
The intervals $I^{K}(l)$ ($K=1,\ldots, \infty$ and  $l=1,\ldots , L$)
are disjoint.

\item
The intervals $I^{K'}(l)$ ($K'=1,\ldots, K$ and $l=1,\ldots , L$)
 and the interval $R^K$
are disjoint.

\item
The intervals $I^{K'}(l)$ ($K'=K+1,\ldots,\infty$ and $l=1,\ldots , L$)
 all belong to the
interval $R^K$.

\item
The union of the intervals $I^{K'}(l)$ ($K'=1,\ldots, K$ and
$l=1,\ldots , L$)
and of $R^K$ is the unit interval:
\beq
\left( \bigcup_{K'=1}^K \bigcup_{l=1}^L I^{K'}(l)\right)
\bigcup R^K = [0,1[
\eeq

\item
The union of the intervals $I^{K+1}(l)$ ($K$ fixed and $l=1,\ldots , L$)
and of $R^{K+1}$ is the
interval $R^K$:
\beq
\left(  \bigcup_{l=1}^L I^{K+1}(l)\right)  \bigcup R^{K+1} = R^K
\eeq

\item
The union of all the intervals $I^{K}(l)$ ($K=1,\ldots,\infty$ and
$l=1,\ldots,L$)
is the unit interval
\beq
\bigcup_{K=1}^\infty \bigcup_{l=1}^L I^{K}(l) = [0,1[
\eeq

\item
The length of the interval $R^K$ is
\beq
\mu (R^K) = 1 - T^{3n/2 + K+1} \leq 2^{-K}
\label{muRK}
\eeq

\item
The length of the union of the intervals $I^{K}(l)$
($K=1,\ldots,\infty$ and
$l$ fixed) is $P(l)$:
\beq
\mu \left( \bigcup_{K=1}^\infty  I^{K}(l) \right)= \sum_{K=1}^\infty
\mu \left(  I^{K}(l)\right )
 = P(l)
\eeq

\end{enumerate}

Bob's strategy is now simple to describe.
Initially, before Alice sends him any information, he chooses a random
number $r$ uniformly distributed in the interval $[0,1[$. He then
carries out the following operations at each round.

{\em Bob's strategy:}
At round $K$, he checks
whether $r$ belongs to $I^{K}(l)$. If so he outputs outcome $l$ and
tells Alice he does not need any more information. On the other hand
if at round $K$, $r$ belongs to
$R^K$ he tells Alice he needs more information.

Because of properties 1 to 5, Bob is sure that at round $K$
$r$ will belong to one of the intervals $I^{K}(l)$ or to $R^K$. Hence
the strategy described above is well defined.
Furthermore, in view of properties 1 and 6, $r$ belongs to one and
only one interval $I^{K}(l)$, hence the protocol will eventually
terminate.

To calculate the probability that Bob outputs outcome $l$, note that
this occurs if and only if $r$ belongs to one of the intervals
$I^{K}(l)$ ($K=1,\ldots,\infty$ and $l$ fixed). The probability
that Bob outputs outcome
$l$ is therefore  equal to
$\mu \left( \bigcup_{K=1}^\infty  I^{K}(l) \right)$. From property 8
this is equal to $P(l)= \langle\Psi | B_l | \Psi \rangle $ as required.

Finally we compute the mean amount of communication required by the
above protocol.
Note that the first round always occurs. The amount of communication
during this round, denoted $C^1$, consists of $(3n/2 +2)2^{n+1}$ bits
of communication sent by Alice to Bob (namely the values of the
coefficients $x_m(j) , y_m(j)$, $j=1,\ldots, 2^n$,
$m=1,\ldots,3n/2+2$) and of one bit of communication sent by Bob to
Alice (telling her whether he could choose an outcome or not).

The subsequent rounds $K\geq 2$ do not always occur. Round $K$ only
occurs if Bob was not able to choose an outcome before round $K$, that
is if $r$ does not belong to any of the intervals $I^{K'}(l)$
($K'=1,\ldots,K-1$ and $l=1,\ldots, L$). Using property 4, this can
be re-expressed as the fact that round $K$ occurs if and only if $r$
belongs to $R^{K-1}$. The probability that round $K$ occurs is
therefore
\beqa
P(\mbox {round K occurs})&=& \mu (R^{K-1})\nonumber\\
&=& 1 - T^{3n/2 + K} \leq
2^{-K+1} \ .
\eeqa
During rounds $K\geq 2$, a certain amount of communication occurs,
always the same,
denoted $C'$. This consists of $2^{n+1}$ bits sent by Alice to
Bob (namely the values of the
coefficients $x_{3n/2+K+1}(j) , y_{3n/2+K+1}(j)$, $j=1,\ldots, 2^n$)
and of one bit of communication sent by Bob to
Alice (telling her whether he could choose an outcome or not).

The average amount of communication is therefore
\beqa
\bar C &=& C^1 + \sum_{K=2}^\infty P(\mbox {round K occurs}) \ C'
\nonumber\\
&\leq& C^1 + C' \sum_{K=2}^\infty 2^{-K-1}
\nonumber\\
&=& C^1 + C' \ .
\eeqa
The average  amount of communication therefore consists of less then
$(3n + 6)2^{n}$
bits sent by Alice to Bob and less then $2$ bits sent by Bob to Alice.
$\Box$

\section{Conclusion}\label{6}

We have
shown in Theorem~\ref{th3} that perfect classical simulation of quantum
communication and entanglement is impossible if the amount of communication
is bounded and the two parties share a finite number of random bits.
Indeed with bounded communication and finite prior shared randomness, only
approximate simulations of quantum communication are possible.

However if we give the parties something slightly more powerful than finite
communication then perfect simulation becomes possible.
One possibility is for Alice and Bob to have an {\em a priori\/} supply of
an infinite number of shared random bits as in \cite{BCT99} (this could, for
instance, be established by having an infinite conversation prior to the
start of the simulation protocol itself).
A second possibility, considered in Theorem~\ref{th5}, is for Alice and Bob
to share no prior randomness and to require that the amount of communication
is only finite on average.
In this case, the amount of communication varies from one simulation to
another, and can sometimes be arbitrarily large.

In all cases, as the number $n$ of qubits or ebits that must be
simulated increases, the amount of classical communication required grows
exponentially with $n$. This has been proven in the bounded
communication scenario in \cite{BCT99}.
We expect this scaling to also hold in the case of simulation without prior
shared randomness and bounded average communication.

{\bf Acknowledgments.}
Part of this work was completed at the 1999 workshop on Complexity,
Computation and the Physics of Information, Isaac Newton Institute,
Cambridge, UK.
We would like to thank Andreas Winter for help with the Appendix.
S.M. and N.C.
acknowledge financial support from the European Science Foundation
and from European Union project EQUIP (contract IST-1999-11063).
S.M. is a research associate of the Belgian National Fund for Scientific
Research.
R.C. is supported in part by Canada's NSERC.
DB is
supported by the U.S. Army Research Office under contract number
DAAG55-98-1-0371.

\appendix

\section{Shared randomness achieved by communication}

In this Appendix, we sketch a proof that the mutual information $I$
between Alice and Bob's outputs (if they share no prior randomness)
is bounded from above by the total number of bits exchanged
in an arbitrary number of rounds of two-way communication.
Assume that, initially, Alice and Bob have each a random
variable denoted respectively as $A_0$ and $B_0$ (this represents
a local source of randomness), but they share no information,
i.~e., $I(A_0{\rm:}B_0)=0$. Then, Alice and Bob communicate via
an arbitrary number of rounds of two-way communication. The first
round consists of Alice sending $B_1$ to Bob, followed by Bob
sending $A_1$ to Alice. So $B_1$ is a function of $A_0$, while
$A_1$ is a function of $B_0$ and $B_1$. In general, the $i$th round
consists in Bob receiving $B_i$ followed by Alice receiving $A_i$.
Again, $B_i$ is a function of $A_0,\cdots A_{i-1}$, and
$A_i$ is a function of $B_0,\cdots B_i$. Assume that this
protocol terminates after $N$ rounds.
Alice then outputs $X=X(A_0,\cdots,A_N)$ which is a funtion of
all the information Alice has, and similarly Bob outputs
$Y=Y(B_0,\cdots,B_N)$.

We first note that the data processing inequality implies that
\begin{equation}
I(X{\rm:}Y) \leq I(A_0, \cdots,A_N{\rm:} B_0,\cdots,B_N) \ .
\end{equation}
 We now bound the right hand side of this equation by
\beqa  \label{entropies}
\lefteqn{ I(A_0,\cdots,A_N{\rm:}B_0,\cdots,B_N)
= I(A_0{\rm:}B_0) } \hspace{1cm} \nonumber\\
&&
+ H(A_1,\cdots,A_N|A_0) +  H(B_1,\cdots,B_N|B_0) \nonumber\\
&&
- H(A_1,\cdots,A_N,B_1,\cdots,B_N|A_0,B_0) \ .
\eeqa
The first term of the rhs of Eq.~(\ref{entropies}) is zero
since there is no initial shared randomness. The second term
of the rhs of Eq.~(\ref{entropies})
measures the amount of randomness received by Alice
during the $N$ rounds in addition to the initial randomness $A_0$.
It is simply bounded from above by the number of bits
of backward communication $C_b$ since
\beqa
 H(A_1,\cdots,A_N|A_0) &\leq&  H(A_1,\cdots,A_N) \nonumber\\
&\leq& H(A_1) + \cdots + H(A_N)\nonumber\\& =& C_b \ .
\eeqa
Similarly, the third term
in the rhs of Eq.~(\ref{entropies}) is bounded from above by the number of bits
of forward communication $C_f$. Finally, using the chain rule for entropies,
the fourth term in the rhs of Eq.~(\ref{entropies}) can be reexpressed as
\beqa
\lefteqn{ H(A_1,B_1|A_0,B_0)+H(A_2,B_2|A_0,A_1,B_0,B_1)+\cdots}
\hspace{1cm} \nonumber\\
&&
+H(A_N,B_N|A_0,\cdots,A_{N-1},B_0,\cdots,B_{N-1})
\eeqa
The $i$th term in this sum can be written as
$H(A_i,B_i|A_0,\cdots A_{i-1},B_0,\cdots B_{i-1})=
H(B_i|A_0,\cdots,A_{i-1},B_0,\cdots,B_{i-1})
+H(A_i|A_0,\cdots,A_{i-1},B_0,\cdots,B_i)$.
These two conditional entropies vanish since $B_i$ depends on
$A_0,\cdots A_{i-1}$, and $A_i$ depends on $B_0,\cdots,B_i$.
Thus, the fourth term in the rhs of Eq.~(\ref{entropies}) is zero.
As a consequence, we have
\beq
I(X{\rm:}Y) \leq C_f + C_b
\eeq
as asserted above.


\end{document}